\documentclass[usenatbib]{mnras}
\usepackage{graphicx, amssymb, aas_macros, natbib, array}
\usepackage{color, colortbl}
\usepackage{multirow}
\usepackage{times}
\newcommand{\vcirc}{V_{\rm{circ}}}

\newcommand{\mvir}{M_{\rm{vir}}}

\newcommand{\mstar}{M_{\star}}
\newcommand{\msun}{M_{\odot}}

\newcommand{\kpc}{{\rm kpc}}

\newcommand{\lcdm}{$\Lambda$CDM}
\newcommand{\mfifty}{M(<\!50\,{\rm kpc})}
\newcommand{\mhundred}{M(<\!100\,{\rm kpc})}

\newcommand{\illustris}{Illustris}

\definecolor{LtGray}{gray}{0.85}
\definecolor{vLtGray}{gray}{0.9}

\title[Mass of the Milky Way from Illustris]
{
The Mass Profile of the Milky Way to the Virial Radius from the Illustris Simulation
}

\author[C. Taylor et al.]
{Corbin Taylor$^1$\thanks{cjtaylor@astro.umd.edu}, 
  Michael Boylan-Kolchin$^2$\thanks{mbk@astro.as.utexas.edu}, 
  Paul Torrey$^{3, 4}$, 
  Mark Vogelsberger$^{3}$, 
  \newauthor and Lars Hernquist$^{5}$
\\
$^1$Department of Astronomy, University of Maryland, 1113 Physical Sciences
Complex (Building 415), College Park, MD 20742-2421, USA\\
$^2$Department of Astronomy, The University of Texas at Austin, 2515
Speedway, Stop C1400, Austin, TX 78712-1205, USA\\
$^{3}$Department of Physics, Kavli Institute for Astrophysics and Space
Research, Massachusetts Institute of Technology, Cambridge, MA 02139, USA\\
$^{4}$TAPIR, Mailcode 350-17, California Institute of Technology, Pasadena, CA
91125, USA\\ 
$^{5}$Harvard-Smithsonian Center for Astrophysics, 60 Garden
Street, Cambridge, MA, 02138, USA}
\begin{document}

 \pagerange{\pageref{firstpage}--\pageref{lastpage}} 
 \pubyear{2016}
 \pagerange{3483--3493}
 \volume{461}
 \maketitle

\label{firstpage}
\begin{abstract}
  We use particle data from the \illustris\ simulation, combined with individual
  kinematic constraints on the mass of the Milky Way (MW) at specific distances
  from the Galactic Centre, to infer the radial distribution of the MW's dark
  matter halo mass. Our method allows us to convert any constraint on the mass
  of the MW within a fixed distance to a full circular velocity profile to the
  MW's virial radius. As primary examples, we take two recent (and discrepant)
  measurements of the total mass within 50 \kpc\ of the Galaxy and find that they
  imply very different mass profiles and stellar masses for the Galaxy. The
  dark-matter-only version of the \illustris\ simulation enables us to compute
  the effects of galaxy formation on such constraints on a halo-by-halo basis;
  on small scales, galaxy formation enhances the density relative to
  dark-matter-only runs, while the total mass density is approximately 20\%
  lower at large Galactocentric distances. We are also able to quantify how
  current and future constraints on the mass of the MW within specific radii
  will be reflected in uncertainties on its virial mass: even a measurement of
  $\mfifty$ with essentially perfect precision still results in a 20\%
  uncertainty on the virial mass of the Galaxy, while a future measurement of
  $\mhundred$ with 10\% errors would result in the same level of uncertainty. We
  expect that our technique will become even more useful as (1) better kinematic
  constraints become available at larger distances and (2) cosmological
  simulations provide even more faithful representations of the observable
  Universe.
\end{abstract}


\section{Introduction}

While living within the Milky Way (MW)  galaxy does have its virtues, easily and
accurately determining the mass distribution of the Galaxy's dark matter halo is
not one of them. This is not for lack of trying, naturally; a variety of
techniques have been crafted for just this purpose, and multiple classes of
kinematic tracers are available. 

The difficulty in measuring the MW's mass distribution is two-fold. First, only
line-of-sight information is available for the vast majority of kinematic
measurements. While great strides are being made in measuring the proper motions
of both individual stars \citep{cunningham2015} and dwarf galaxies (e.g.,
\citealt{Piatek2007, Sohn2013, vanDerMarel2014, pryor2015}) at large
Galactocentric distances in the MW's halo, the number of tracers at
$\sim 50-100$ kpc with full 6D phase space information will remain small even in
the Gaia era \citep{deBruijne2014}.  Perhaps more importantly, the level of
precision desired for the MW's mass is simply higher than is the case for other
galaxies. Whereas a factor of $\pm2$ uncertainty in the mass of a typical
galaxy's halo would be considered an excellent measurement, it is often thought
of more as an embarrassment in the case of the MW.

For example, if we take a dark matter halo mass of $10^{12}\,\msun$ as a
fiducial estimate for the MW, changes by a factor of 2 in either direction are
the difference between: (1) an implied conversion efficiency of baryons into
stars of $\approx 70\%$ (at $M=5\times 10^{11}\,\msun$) and $16\%$ (at
$2 \times 10^{12}\,\msun$); (2) eliminating the too-big-to-fail problem
\citep{Boylan-Kolchin2011a, Boylan-Kolchin2012} and severely exacerbating it
\citep{Wang2012, Vera-Ciro2013, Jiang2015}; and (3) placing the Large Magellanic
Cloud and the Leo I dwarf spheroidal on unbound versus bound orbits
\citep{kallivayalil2006, kallivayalil2013, besla2007, Boylan-Kolchin2013}. Our
understanding of the MW in cosmological context relies on our ability to know
its mass to high precision.

While uncertainties are most pronounced in the outer dark matter halo of the MW,
where there are few tracers of the total mass, they also persist at small
Galactocentric distances: there are disagreements about the mass within the solar
circle at the 25\% level (e.g., \citealt{Bovy2012, schonrich2012}). At 40--80
kpc, estimates differ at the 50\% level (see, e.g., \citealt{Williams2015}).

In this paper, we take an alternate approach to constraining the mass
distribution of the MW. Cosmological hydrodynamic simulations are now
producing galaxies that match a variety of observations both for statistical
samples of galaxies and for individual galaxies themselves. In particular, both
the \illustris\ \citep{Vogelsberger2014a} and Eagle \citep{Schaye2015}
simulations use $\sim 10^{10}$ particles within $\sim 100$ Mpc boxes, meaning
they contain thousands of haloes with masses comparable to that of the MW, each
with of the order 1 million particles within the virial radius. The successes of
these models, and the underlying successes of the $\Lambda$ cold dark matter (\lcdm) model, motivate using
the results of cosmological simulations to constrain the mass distribution of
the MW.

There are a number of ways one could use cosmological simulations for this
purpose. Indeed, several previous works on the mass of the MW have used
cosmological simulations in some capacity. One possibility is to use dark matter
haloes from large cosmological simulations as point particles and calibrate the
timing argument \citep{Kahn1959} for measuring the total mass of the MW
\citep{Li2008, Gonzalez2014}. Alternately, properties of satellites from
cosmological simulations can be compared to those of MW satellites such
as the Magellanic Clouds, yielding estimates of the virial mass of the MW
\citep{Boylan-Kolchin2011b, Busha2011, Gonzalez2013, Fattahi2016}. Yet another possibility is
to use individual, high-resolution simulations of Milky Way-sized haloes in
conjunction with kinematic information about dwarf satellites of the MW
\citep{Boylan-Kolchin2013, Barber2014}. Cosmological hydrodynamic
simulations of individual MW-mass haloes have also been used to calibrate
kinematic analyses of tracer populations in order to measure the mass of the MW
\citep{Xue2008, Rashkov2013, Piffl2014, Wang2015}.

Our approach is to use importance sampling in a homogeneously-resolved,
large-volume cosmological simulation, weighing each simulated halo by its level of consistency with the MW; for a clear description of this technique
applied to cosmological simulations, see \cite{Busha2011}. By taking any
individual constraint and using it to perform importance sampling from
simulations, we can find the mass distributions of haloes that are consistent
with the imposed constraint. An advantage of this technique is that it allows us
to easily map different constraints, with different errors, on to mass
distributions for the MW and its dark matter halo.

Variants of importance sampling have been used to measure the mass of the MW
\citep{Li2008, Boylan-Kolchin2011b, Busha2011, Gonzalez2014}. However, previous work has
generally focused on using dark-matter-only (DMO) simulations to measure the
total (virial) mass of the MW. With hydrodynamic simulations, we are able to
make two improvements. First, we are able to measure the mass
\textit{distribution} of the MW in simulations that self-consistently model the
effects of galaxy formation on the dark matter haloes of galaxies. Secondly, we are
able to compare our constraints directly to those obtained from DMO simulations,
as a DMO version of \illustris\ is also publicly available. By matching objects
between the two simulations, we are able to investigate, in detail, the effects
of baryonic physics on inferences of the mass distribution of the MW from
cosmological simulations.

We generally use the mass within 50 kpc as our primary constraint, as this is
approximately the largest radius where stellar kinematic tracers are found in
large enough numbers to facilitate a mass measurement. We also provide estimates
for how a measurement of the mass within 100 kpc -- which future surveys may
provide -- will improve our knowledge of the mass distribution at even larger
radii.

This paper is structured as follows. Section~\ref{sec:methods} describes our
basic approach, provides information about the Illustris simulation, and
describes our primary analysis of the simulation. Section~\ref{sec:results}
contains our main results regarding the mass distribution of the MW as
derived from haloes taken from the \illustris\ simulation. We also quantify how
inferences on the enclosed mass at large scales (at 250 kpc and various spherical
overdensity values) depend on the measured mass within 50 kpc and quantify the
stellar masses of galaxies having haloes consistent with the adopted mass
constraint. A discussion of our results and prospects for future improvements is
given in Section~\ref{sec:discussion}; our primary conclusions are given in
Section~\ref{sec:conclusions}. Throughout this paper, error bars give 68\%
confidence intervals unless otherwise noted.

\section{Methods}
\label{sec:methods}

\subsection{Simulations and Importance Sampling}
\label{sec:simulations}

Our analysis is based on the \illustris\ suite of cosmological simulations
\citep{Vogelsberger2014a}, which consists of paired hydrodynamic and
DMO simulations at three different resolution levels. Each
simulation uses a periodic box of length $75\,h^{-1}$ Mpc and an initial
redshift of $z=127$. The highest resolution simulation, \illustris-1, uses
$1820^3$ dark matter particles and an equal number of hydrodynamic cells
initially, with a spatial resolution of $1\,h^{-1}$ kpc for the dark matter. The
DMO version of this simulation, \illustris-Dark-1, uses identical initial
conditions but treats the baryonic component as collisionless mater. Two lower
resolution simulation of the same volume, \illustris-2 and \illustris-3, were
also performed, with 8 and 64 times fewer particles, respectively. The
background cosmology for all of the simulations was chosen to be consistent with
Wilkinson Microwave Anisotropy Probe-9 results \citep{Hinshaw2013}: $\Omega_{\rm m,0} = 0.2726$,
$\Omega_{\Lambda, 0} = 0.7274$, $\Omega_{\rm b,0} = 0.0456$, $\sigma_{8} = 0.809$,
$n_{\rm s} = 0.963$, and $h = 0.704$. Haloes and subhaloes in the \illustris\ simulations were identified using a friends-of-friends algorithm followed by SUBFIND \citep{Springel2001}. For further information about the
\illustris\ suite,\footnote{The \illustris\ data are all publicly available (http://www.illustris-project.org/); see
  \citet{Nelson2015} for further information.} including details about the
implementation of galaxy formation physics, see \citet{Vogelsberger2013,Vogelsberger2014b}.

Using Illustris to inform our understanding of the mass distribution of the
MW requires calculations of the mass profiles of an unbiased sample of
dark matter haloes within the simulation. Although the halo catalogues provide the
centres for each halo (we only consider central halos, not subhaloes, as possible centres), a brute-force calculation of the mass profile for each
halo is prohibitively expensive, as it requires repeated searches through the
$\sim 10^{10}$ particles of the simulation. We instead use a K-D tree algorithm,
taking into account the periodic boundary conditions of the simulation
volume. The algorithm was verified against brute-force calculations applied to
Illustris-3 and Illustris-2.

In addition to considering the mass within spherical apertures, we also compute
spherical overdensity masses with respect to three common overdensity choices:
$M_{\rm 200,c}$ (measured with respect to $200\,\rho_{\rm crit}$),
$M_{\rm 200,m}$ (measured with respect to
$200\,\rho_{\rm m}\approx 55\,\rho_{\rm crit}$ for the Illustris cosmology at
$z=0$), and $M_{\rm vir}$ (measured with respect to
$\Delta_{\rm vir}\,\rho_{\rm crit}$; for the Illustris cosmology at $z=0$,
$\Delta_{\rm vir} \approx 97$; \citealt{Bryan1998}).

\begin{figure*}
\centering
\includegraphics[width=0.48\linewidth]{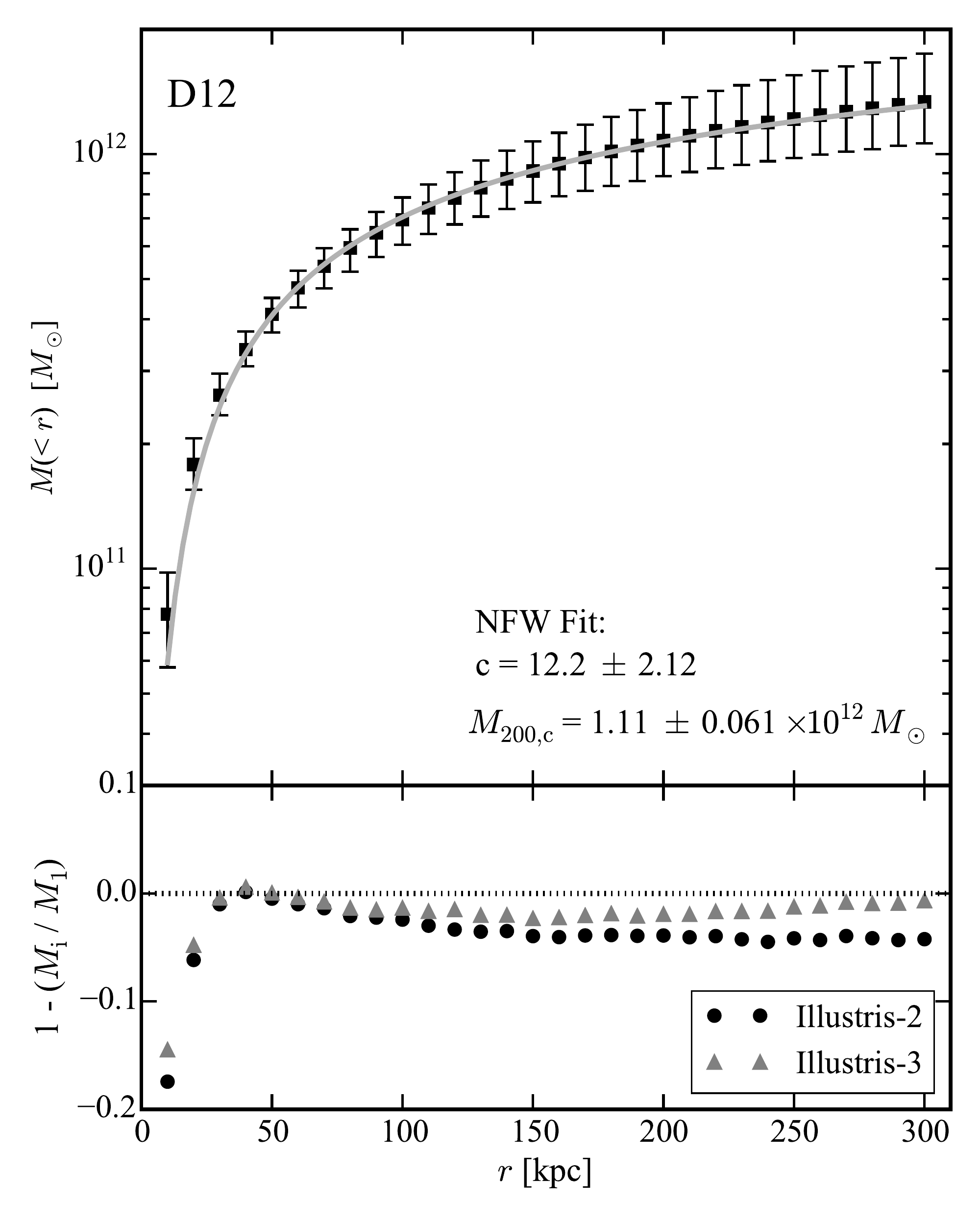}
\includegraphics[width=0.48\linewidth]{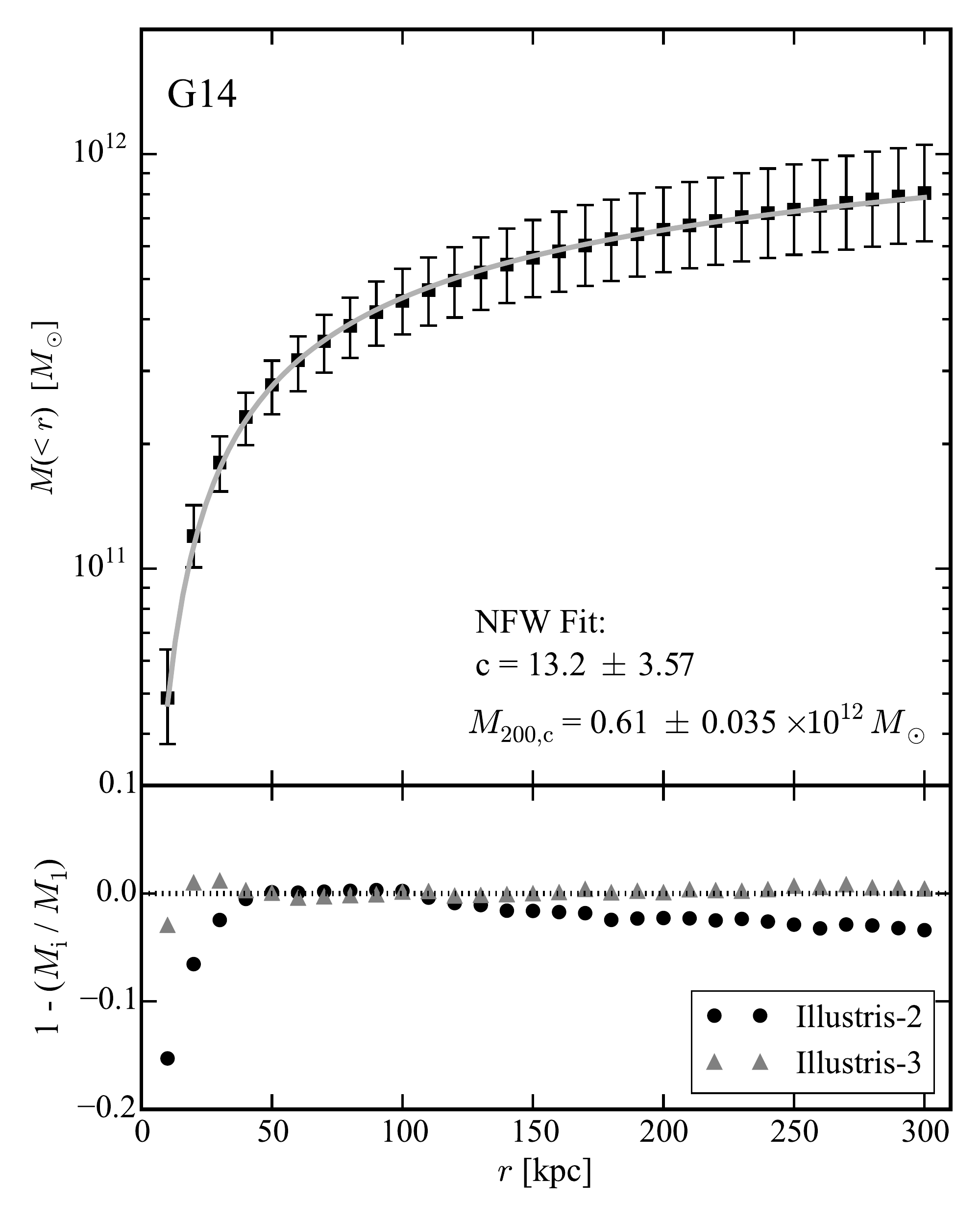}
\caption{Top: the mass distribution $M(<r)$ derived from \illustris-1,
  using the D12 (left) and G14 (right) constraints on $\mfifty$. Error bars
  represent 68\% confidence intervals. The grey lines show best-fitting NFW
  profiles for the full mass distribution (dark matter and baryonic); the NFW
  fit parameters are given in the figure. The two constraints result in very
  different estimates of $M_{\rm 200,c}$; the concentration parameters are less
  disparate. Bottom: residuals between the mass distribution obtained
  from \illustris-1 and \illustris-2 (black circles) or \illustris-3 (grey
  triangles). At $r<30$ kpc, systematic differences are evident; these likely
  result from a combination of numerical resolution and differences in the
  stellar masses at fixed halo mass. Small differences of 2-5\% exist at larger radii; however, such deviations are much smaller than the uncertainties we derive in Table \ref{table:confidenceIntervals}.}
\label{mInR}
\end{figure*}

\subsection{Statistical Analysis}
Our basic framework is to consider the Illustris simulation a plausible model of
galaxies in our Universe, then to assign each halo in the simulation a weight
based on how closely its enclosed mass at some radius\footnote{Here and throughout this work, we use 'mass' to refer to the enclosed mass (as opposed to mass within a spherical shell)} (we typically use 50 kpc in what
follows) matches observational data. The resulting weights for the halo sample
then provide a constraint on the enclosed mass of the MW at other radii.

In more detail, we take an observational measurement of the total MW mass
within a specific radius and assign a weight to each halo in the \illustris\
galaxy catalog: assuming the observed mass has a value of $\mu$ and an
associated (Gaussian) error of $\sigma$, then the weight $W_i$ contributed by an
individual halo $i$ with enclosed mass $M_i$ at the specified radius is
\begin{equation}
\centering
W_i = \frac{1}{\sqrt{2\rm{\pi}}\sigma}
\exp\left(\frac{-(M_i-\mu)^{2}}{2\sigma^{2}}\right)\,.
\label{eq:singleprob}
\end{equation}
We can then construct the full mass or circular velocity profile and compute the
total stellar or halo mass that is consistent with the observed constraint by
using the distribution of weights assigned to the haloes. In this analysis, we
assume that observed constraints all follow Gaussian distributions, consistent
with the analyses we incorporate, but this technique can be easily extended to
any other analytic or numerical probability distribution. In what follows, we
quote median values and confidence intervals that are centred on the median and
contain 68\% of the probability distribution.

The primary observational constraint we use is the total mass of the MW within
50 kpc, $\mfifty$. There are many literature estimates of the MW's mass at
approximately this scale (e.g., \citealt{Wilkinson1999, battaglia2005, Xue2008,
  brown2010, gnedin2010, Kafle2014, Eadie2015}), in large part because (1) this
is approximately the distance to which large samples of blue horizontal branch
(BHB) stars are currently available from surveys such as the Sloan Digital Sky Survey, and (2) the LMC
lies at a Galactocentric distance of $\approx 50\,\kpc$, meaning estimates of
the MW mass based on LMC's dynamics directly constrain $\mfifty$. We focus on
two recent and disparate measurements of $\mfifty$: \citet[hereafter D12]{deason2012}, who used
BHB stars and found $\mfifty=4.2 \pm 0.4 \times 10^{11}\,\msun$, and
\citet[hereafter G14]{gibbons2014}, who used the Sagittarius stream to measure
$\mfifty=2.9 \pm 0.4 \times 10^{11}\,\msun$ \citep{Gomez2015}.  These measurements are clearly
incompatible at the $3\,\sigma$ level and therefore are useful for showing the
effects of varying $\mfifty$ on the inferred mass distribution at larger
radii. In Sec.~\ref{subsec:mass_at_50}, we explicitly show how estimates of
$M_{\rm 200,c}$ vary as a function of $\mfifty$.

In principle, a complete analysis would include every dark matter halo in
Illustris. In practice, however, only a relatively narrow range of masses
contribute any weight to our inferences. We therefore restrict our analysis to all
haloes with $M_{\rm 200,c} = (0.1-10)\times10^{12}\msun$, which includes $14\,192$
haloes for Illustris-1, $14\,316$ haloes for Illustris-2, and $12\,885$ haloes for
Illustris-3.  As we show below, this mass range is more than sufficient for
including all relevant haloes in our analysis and does not bias our results in
any way. 

\section{The Mass Distribution of the MW}
\label{sec:results}

\subsection{The MW's radial mass profile}
\label{subsec:minr}

\begin{figure*}
\centering
\includegraphics[width=0.495\linewidth]{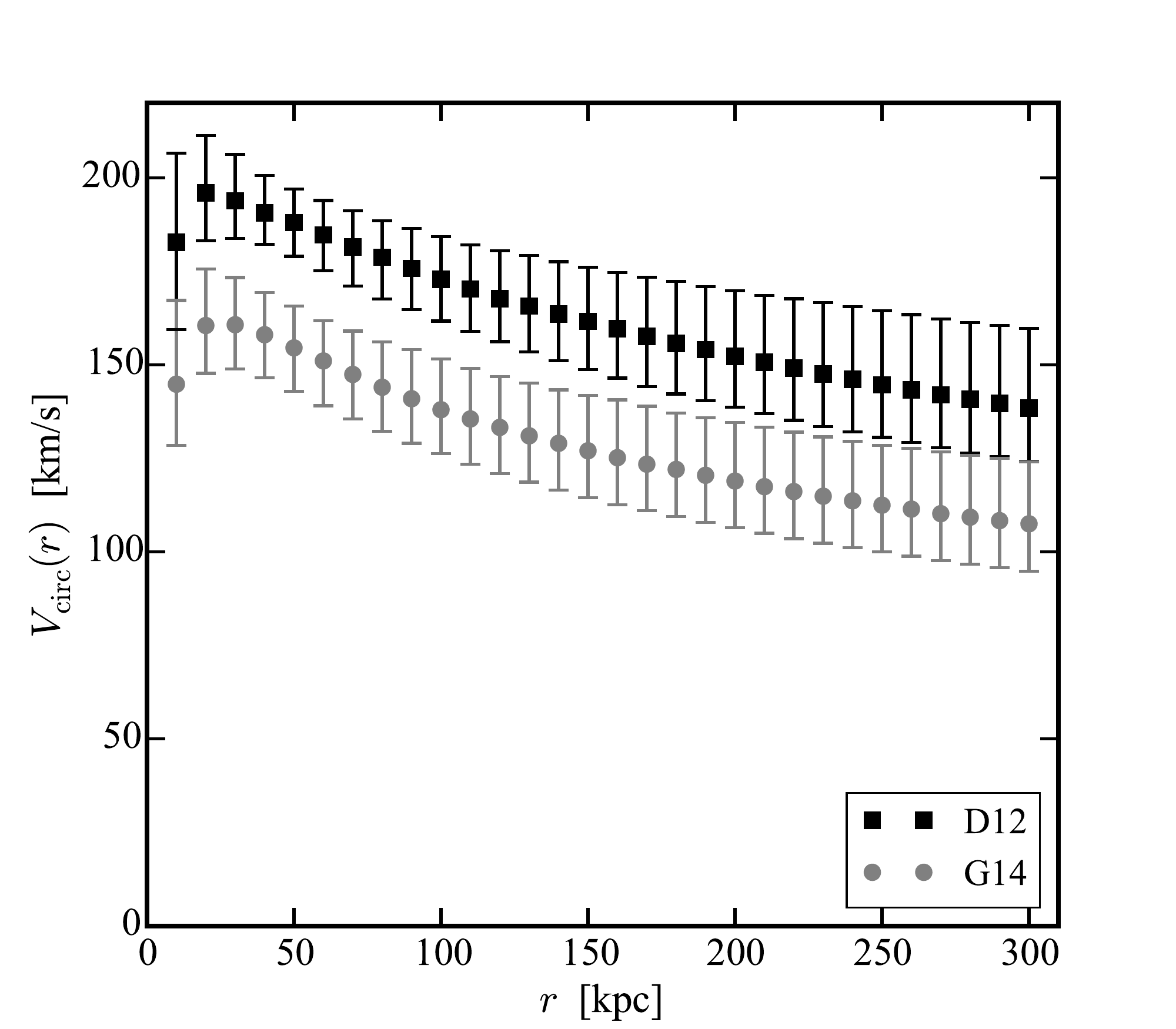}
\includegraphics[width=0.495\linewidth]{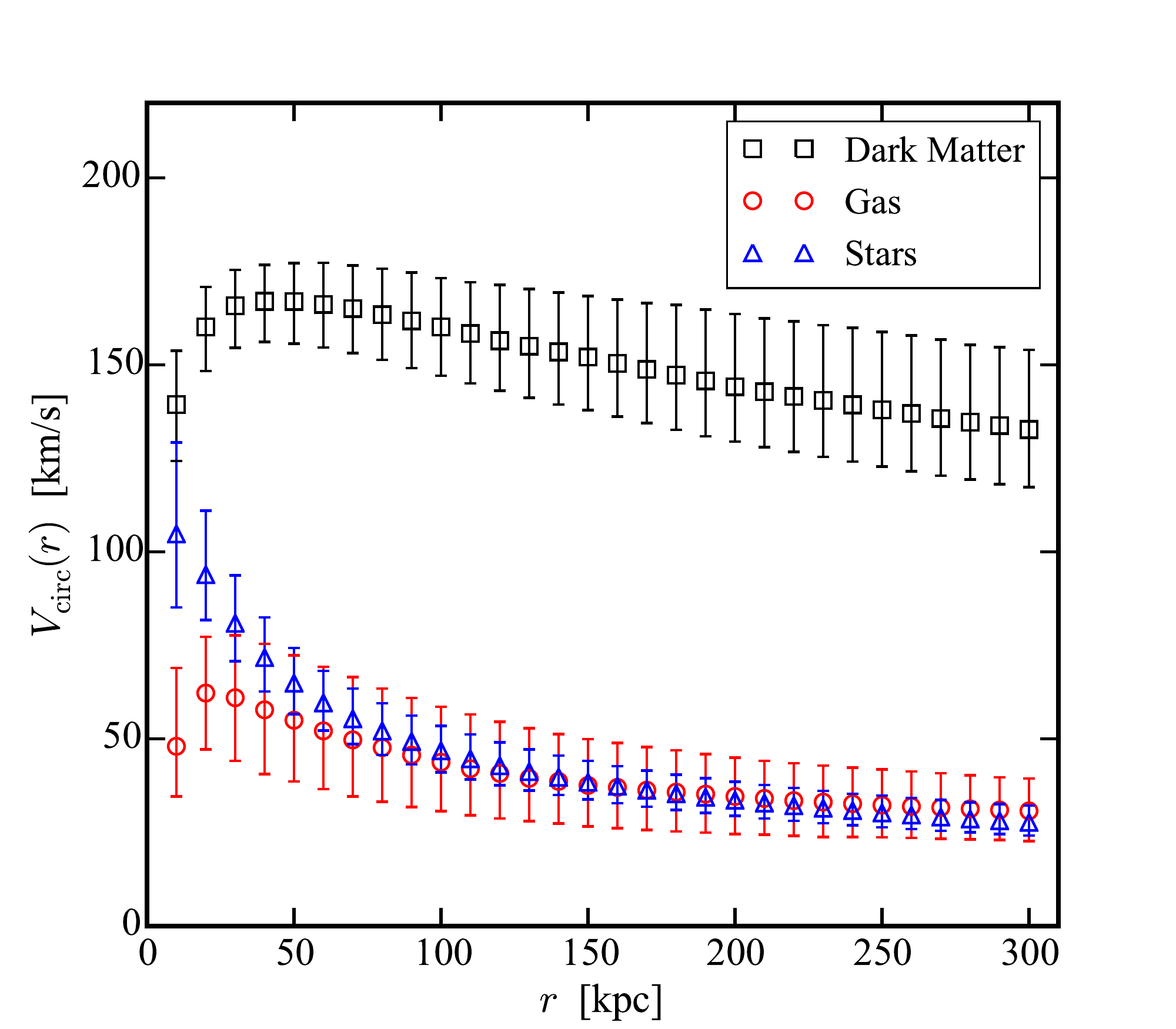}
\caption{Circular velocity curves. Left: $\vcirc(r)$ for the mass
  profiles given in Fig.~\ref{mInR}. The overall mass profile using G14's
  constraint is lower at every Galactocentric distance compared to the profile
  derived using the D12 constraint, and the 68\% confidence intervals are
  disjoint up to $280\,\kpc$.  Right: a decomposition of the
  circular velocity profile derived using the D12 constraint (black points in
  the left panel of Fig.~\ref{mInR}) into separate contributions from dark
  matter (black), stars (blue) and gas (red). At all radii probed here, dark
  matter dominates. The contribution from gas matches that from stars near a
  halo-centric distance of 100 kpc.}
\label{vcirc}
\end{figure*}

Fig.~\ref{mInR} presents the mass distributions obtained using the constraints
on $\mfifty$ from D12 (left-hand panel) and G14 (right-hand panel) from the Illustris-1
sample. The best-fitting Navarro-Frenk-White (\citeyear[hereafter,
NFW]{Navarro1997}) profiles for the \textit{total} mass distribution are given in
the figure as well. The fits were performed over the radial range of 40--300
kpc, as we find a lack of convergence among different resolution versions of
\illustris\ on smaller scales (see below; convergence in density profiles should
occur at smaller scales, as density is a differential quantity while mass and
circular velocity are cumulative quantities). Unsurprisingly, given the
significantly higher value of $\mfifty$ found in D12 relative to G14, the
best-fitting NFW value of $M_{\rm 200,c}$ for D12 is much larger than for G14,
$1.1\times10^{12}\,\msun$ versus $0.61\times 10^{12}\,\msun$. The best fitting
concentration parameters are similar: $c_{\rm 200,c}=12.2\pm2.12$ for D12 and
$c_{\rm 200,c}=13.2\pm 3.57$ for G14. Both of these concentrations are larger
than those derived from large DMO simulations, which typically find
$c_{\rm 200,c}\approx 8.33$ for haloes of $M_{\rm200,c} \approx 10^{12}\,\msun$
(e.g., \citealt{dutton2014}).

The lower panels of the figures show the fractional differences of Illustris-2
and Illustris-3 with respect to their high-resolution counterpart, with error
bars representing 68\% confidence intervals. There are relatively large
differences between the different levels of resolution at relatively small radii
($r<30$ kpc), while differences are much less substantial farther away from
Galactic Centre.  With a gravitational softening length $\sim$ 1 kpc and
baryonic sub-grid routines tailored specifically to the highest resolution
simulation. This lack of convergence on small scales is not
surprising. For instance, \citet{Schaller2016} show that the dark matter
density profiles of Eagle galaxy haloes are only converged at $\approx 20$ kpc
(their fig. 3). We therefore strongly caution against extrapolating the
  NFW fits presented in this paper to small radii ($r \la 30$ kpc). If future generations of simulations provide well-converged results at smaller radii, the dark matter fraction within $\sim 2$ disc scale lengths will likely provide important constraints on feedback models \citep{Courteau2015}.

The circular velocity profiles, $V_{\rm circ}(r)$, corresponding to the
cumulative mass profiles of Fig.~\ref{mInR} are shown in the left-hand panel of
Fig.~\ref{vcirc}. This highlights the large difference in the two
determinations of the MW potential, as well as how this difference persists in
predicted profiles out to 300 kpc. It is only at distances $>250$ kpc that the
68\% confidence intervals begin to overlap.

The distribution of mass among dark matter, stars, and gas within any given radius
is interesting to consider: observationally, we can measure the stellar mass
with reasonable accuracy and infer the dark matter mass, but constraining the
distribution of the Galaxy's gaseous component at large distances is much more
difficult (see, e.g., \citealt{Gupta2012, Fang2013}). In the right-hand panel of
Fig.~\ref{vcirc}, we plot the circular velocity profile decomposed into the
contributions from each of these components. Dark matter dominates the potential
at all radii we study, and while stars substantially outweigh the gas for
$r \la 50$ kpc, the two contribute approximately the same mass by
$r \approx 100$ kpc.

\subsection{Mass constraints within specific radii}
\label{subsec:mass_at_50}
In this subsection, we explore predictions for enclosed masses at specific radii in more
detail. In particular, we are interested in understanding how observational
constraints at $\mfifty$ translate into inferences on masses at other radii. We
consider both individual physical radii (in particular, 100 and 250 kpc) and
various definitions of spherical overdensity masses ($M_{\rm 200,c}$, $M_{\rm
  vir}$, and $M_{\rm 200,m}$).

\begin{figure}
\centering
\includegraphics[width=\linewidth]{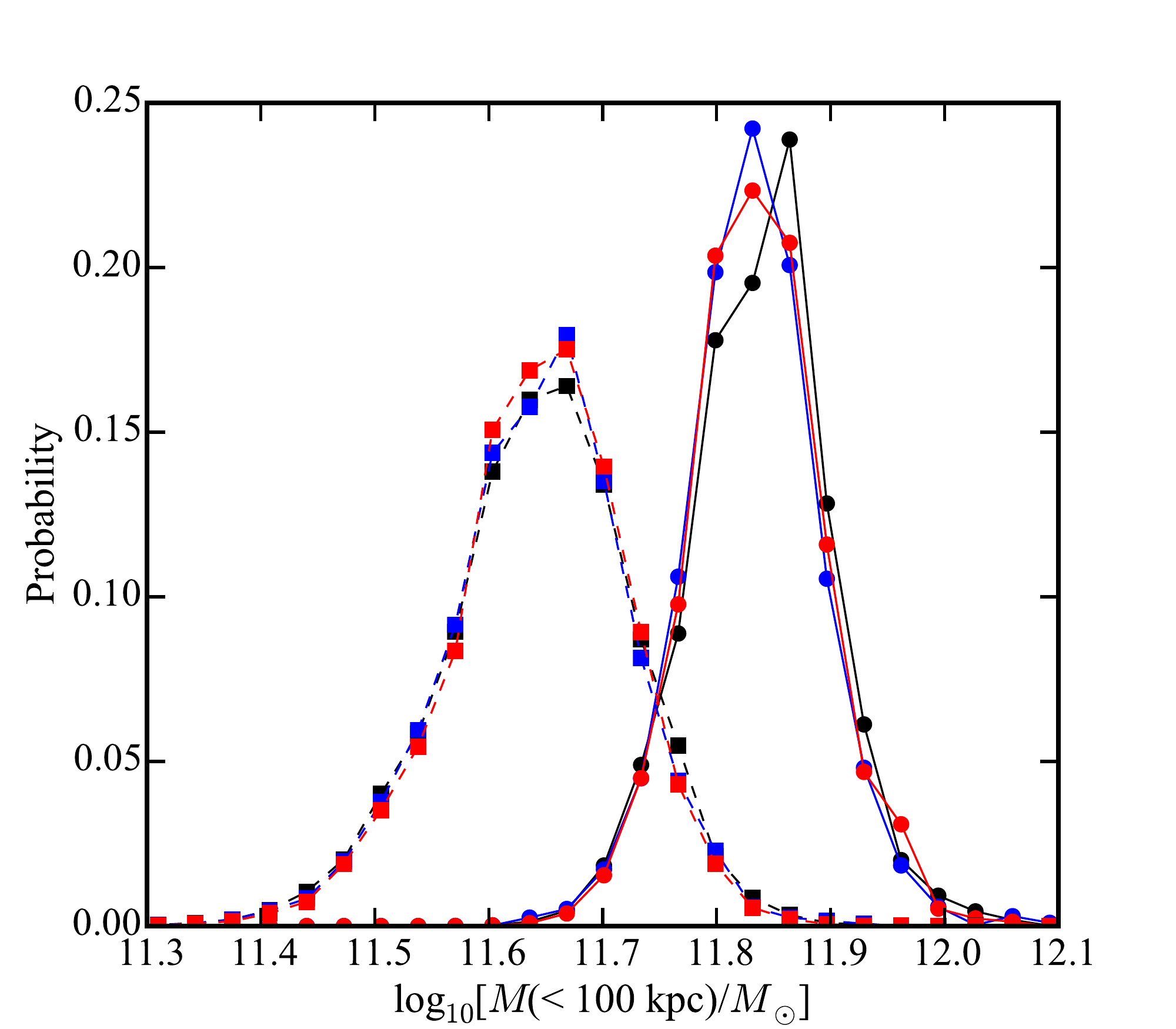}
\caption{The probability distribution of $\mhundred$ derived from the G14
  (squares, 
  connected by dashed lines) and D12 (circles, connected by solid lines)
  constraints on $\mfifty$. The colours represent the individual resolution
  levels: Illustris-1 (black), Illustris-2 (blue), and Illustris-3 (red). The
  excellent agreement across the three levels of resolution indicates that the
  total mass profiles are well-converged in Illustris.}
\label{m100-m50}
\end{figure}

Fig.~\ref{m100-m50} presents the probability distribution for $\mhundred$,
with black, blue, and red symbols representing Illustris-1, Illustris-2, and
Illustris-3 respectively.  The results using the D12 constraint on $\mfifty$ are
presented as circles connected with solid lines, while those using the G14
constraint are shown as squares with dashed connecting lines. As expected, and
shown previously, the D12 constraint on $\mfifty$ results in a significantly higher
predicted total mass within 100 kpc (approximately 0.2 dex). The smaller
(relative) error quoted in D12 also results in a narrower distribution for
$\mhundred$.

Perhaps the most important aspect of Fig.~\ref{m100-m50} is the
\textit{excellent} convergence seen across the three Illustris simulations (a
factor of 64 in mass resolution and 4 in force resolution). Not only is the peak
or median value well converged, the entire distribution is essentially identical
in each case. This indicates that, while masses on small scales (10--30 kpc) are
affected by resolution and baryonic physics, enclosed masses at larger radii are not
subject to such effects. The consistency of the mass distributions at large radii,
subject to a constraint at 50 kpc, points to robustness of our technique for
constraining the mass distribution of the MW.

\begin{table}
\setlength{\extrarowheight}{5pt}
  \caption{Median values, along with 68\% and 90\% confidence intervals, for
    mass measures explored in this paper; all masses are expressed in 
    units of $10^{12}\,\msun$.
    In each case, we calculate values using constraints from both D12 (column 2) and 
    G14 (column 3) on each of the three 
    \illustris\ resolution levels. Good convergence across the three levels of
    resolution is evident.}
\resizebox{\linewidth}{!}{
\begin{tabular}{ l| l l}
  \hline
  & D12 & G14\\
  \hline 
  \hline 
  & \multicolumn{2}{c}{Illustris-1}\\
  $M_{\rm 200,c}$ & $1.12^{+0.370\,(0.747)}_{-0.240\,(0.357)}$ & $0.612^{+0.196\,(0.384)}_{-0.148\,(0.227)}$ \\
  $M_{\rm vir}$ & $1.30^{+0.511\,(1.12)}_{-0.304\,(0.445)}$ & $0.711^{+0.251\,(0.522)}_{-0.179\,(0.274)}$ \\
  $M_{\rm 200,m}$ & $1.48^{+0.642\,(1.49)}_{-0.361\,(0.536)}$ & $0.798^{+0.306\,(0.632)}_{-0.213\,(0.319)}$ \\
  $M(< 100\, \kpc)$ & $0.695^{+0.091\,(0.166)}_{-0.090\,(0.149)}$  & $0.443^{+0.087\,(0.148)}_{-0.076\,(0.127)}$ \\
  $M(< 250\, \kpc)$ & $1.22^{+0.334\,(0.631)}_{-0.236\,(0.355)}$ & $0.736^{+0.209\,(0.406)}_{-0.164\,(0.255)}$\\
  \hline
  & \multicolumn{2}{c}{Illustris-2}\\
  $M_{\rm 200,c}$ & $1.06^{+0.296\,(0.674)}_{-0.196\,(0.297)}$ & $0.597^{+0.166\,(0.340)}_{-0.134\,(0.206)}$ \\
  $M_{\rm vir}$ & $1.24^{+0.419\,(0.954)}_{-0.243\,(0.366)}$ & $0.691^{+0.213\,(0.430)}_{-0.163\,(0.250)}$ \\
  $M_{\rm 200,m}$ & $1.40^{+0.490\,(1.19)}_{-0.306\,(0.442)}$ & $0.766^{+0.253\,(0.543)}_{-0.186\,(0.283)}$ \\
  $M(< 100\, \kpc)$ & $0.678^{+0.090\,(0.164)}_{-0.078\,(0.129)}$ & $0.444^{+0.077\,(0.144)}_{-0.077\,(0.124)}$ \\
  $M(< 250\, \kpc)$ & $1.17^{+0.283\,(0.588)}_{-0.194\,(0.293)}$ & $0.714^{+0.184\,(0.360)}_{-0.148\,(0.229)}$\\
  \hline
  & \multicolumn{2}{c}{Illustris-3}\\
  $M_{\rm 200,c}$ & $1.09^{+0.332\,(0.626)}_{-0.193\,(0.304)}$  & $0.614^{+0.170\,(0.332)}_{-0.135\,(0.214)}$\\ 
  $M_{\rm vir}$ & $1.29^{+0.441\,(0.853)}_{-0.249\,(0.386)}$ & $0.712^{+0.220\,(0.434)}_{-0.160\,(0.253)}$ \\ 
  $M_{\rm 200,m}$ & $1.46^{+0.562\,(1.252)}_{-0.293\,(0.450)}$ & $0.804^{+0.256\,(0.519)}_{-0.190\,(0.298)}$ \\ 
  $M(< 100\, \kpc)$ & $0.685^{+0.092\,(0.183)}_{-0.080\,(0.134)}$ & $0.444^{+0.078\,(0.133)}_{-0.073\,(0.122)}$ \\ 
  $M(< 250\, \kpc)$ & $1.20^{+0.300\,(0.546)}_{-0.192\,(0.302)}$ & $0.741^{+0.178\,(0.342)}_{-0.148\,(0.238)}$ \\ 
  \hline
  \hline
\end{tabular}
}

\label{table:confidenceIntervals}
\end{table}

Inferred values of aperture masses within 100 and 250 kpc and three different
spherical overdensity masses, along with 68\% and 90\% confidence intervals, are
given in Table~\ref{table:confidenceIntervals}. The estimated virial mass,
$\mvir$, using D12 is $1.3 \times 10^{12}\,\msun$, with a 90\% confidence
interval of $0.86-2.3\times 10^{12}\,\msun$. This is similar to the result of
\citet{Boylan-Kolchin2013}, who found a 90\% confidence interval of
$1.0-2.4\times10^{12}\,\msun$ for $\mvir$ based on the dynamics of the Leo I
satellite galaxy. Using the G14 estimate of $\mfifty$, we find a median value of
$\mvir=0.71 \times 10^{12}\,\msun$ with a 90\% confidence interval of
$0.44-1.2 \times 10^{12}\,\msun$, both of which are substantially lower than our
inference based on the results of D12. These results highlight the importance of
accurate determinations of $\mfifty$ for understanding the large-scale
properties of the MW. We note that the 99.95\% confidence interval for
haloes consistent with the D12 constraint is
$5.17\times 10^{11}< M_{\rm 200,c}<5.06\times10^{12}\,\msun$ (the range for the
G14 constraint is $2.12\times 10^{11}< M_{\rm 200,c}<3.48\times10^{12}\,\msun$),
confirming that our range of $10^{11} \le M_{\rm 200, c} \le 10^{13}\,\msun$ is
more than sufficient for inferences about the mass of the MW.  

\begin{figure}
\centering
\includegraphics[width=\linewidth]{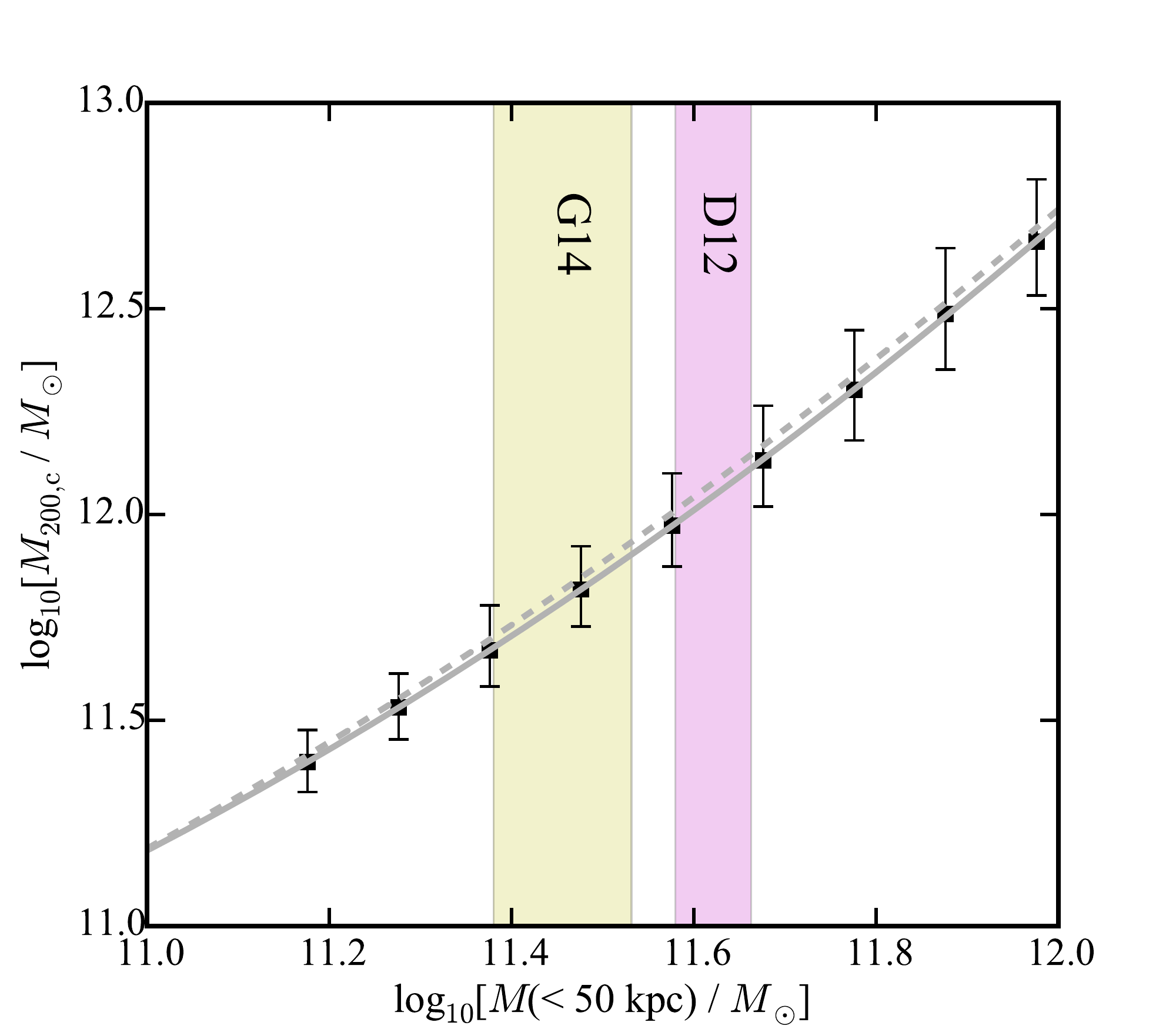}
\caption{The dependence of the inferred value of $M_{\rm 200,c}$ on the input
  (measured) value of $\mfifty$. The data points with error bars show values of
  $M_{\rm 200,c}$ based on our weighting procedure, assuming a 10\% error in
  $\mfifty$. The G14 and D12 determinations of $\mfifty$ are highlighted with
  yellow and magenta vertical bands, respectively, with the widths of the bands
  showing the 68\% confidence intervals. The best-fitting log-quadratic relation
  between $\mfifty$ and $M_{\rm 200,c}$ (given in equation~\ref{eq:mfifty_fit}) is
  plotted as a solid grey line, while the dashed grey line shows the fit to the
  unweighted data; see the text for details. This relation can be used to map any
  constraint on $\mfifty$ to an inferred value of $M_{\rm 200,c}$.}
\label{mcrit-function}
\end{figure}

 Given the uncertainties in $\mfifty$, it is also important to understand how
inferences of spherical overdensity masses depend on $\mfifty$. To do this, we
assume that $\mfifty$ can be measured with an accuracy of 10\% (i.e.,
$X \pm 0.1\,X$) and compute the resulting median value and 68\% confidence
intervals for $M_{\rm 200,c}$. The resulting dependence of $M_{\rm 200,c}$ on
$\mfifty$ is shown in Fig.~\ref{mcrit-function}, where the error bars show
68\% confidence intervals. It is clear that there is a strong correlation
between $\mfifty$ and $M_{\rm 200,c}$. We fit this with a quadratic function in
log space:
\begin{eqnarray}
  &&\log_{10}\left(\frac{M_{\rm 200,c}}{\msun}\right) = A +B\,\mu+C\,\mu^2\,,\label{eq:mfifty_fit}
  \\
  &&\mu = \log_{10}\left(\frac{M(<50\,{\rm kpc})}{4\times10^{11}\,\msun}\right)\,. \nonumber 
\end{eqnarray}
Fitting to the weighted results plotted in Fig.~\ref{mcrit-function}, we find
$A=12.0, \,B=1.60, \,C=0.373$ with an rms scatter of 0.069, whereas fitting the
unweighted data, we find $A=12.0, \,B=1.62, \,C=0.325$ with an rms scatter of
0.067. The latter is offset slightly higher at fixed $\mfifty$, as the weighted
results naturally involve averaging over the dark halo mass function within each bin, which is a steeply declining function of mass, whereas the unweighted results do not.

Equation~\ref{eq:mfifty_fit} can be used to convert any constraint on $\mfifty$
into a constraint on $M_{\rm 200,c}$. It is also straightforward to convert this
fit to a constraint on $M_{\rm vir}$ or $M_{\rm 200,m}$, as $\mvir \approx
1.17\, M_{\rm 200,c}$ and $M_{\rm 200,m}\approx 1.32\,M_{\rm 200,c}$ for the
typical mass profiles in \illustris. If Equation~\ref{eq:mfifty_fit} or a similar relation holds broadly for other
hydrodynamic simulations with different galaxy formation physics
implementations, then it will be of tremendous value for MW mass inference
studies. We plan to examine this issue in more detail in future work (and see
further discussion below).

\subsection{The Impact of Baryonic Physics}
\label{subsec:baryonic_physics}
Our primary analysis, presented over the previous subsections, makes use of the
highest resolution \illustris\ simulation. This, and all other hydrodynamic
simulations of the evolution of a representative galaxy population over cosmic
time, require a number of assumptions in order to produce a realistic set of
galaxies. One of the primary calibrations for \illustris, for example, was to
match the $z=0$ galaxy stellar mass function. As shown in fig. 7 of
\cite{Vogelsberger2014b}, the galaxy formation prescriptions in \illustris\
result in notable changes in the total masses of dark matter haloes over a wide
range in halo mass. Moreover, these changes depend on specific choices made in
the galaxy formation modelling, as the galaxy formation modelling within the Eagle
simulation results in substantially different effects on halo masses (see
fig.~1 of \citealt{Schaller2015}).

It is not a priori obvious whether using the DMO run should result in
similar or different predictions from the fully hydrodynamic simulation, and if
the results are different, it is not clear whether they will be higher or
lower. Certainly, we expect that the formation of a galaxy will lead to a more
centrally concentrated mass distribution relative to the DMO run, to some
extent. Adiabatic contraction of the dark matter in response to gas cooling will
also tend to increase the amount of dark matter in the central regions of the
halo. On the other hand, it is well established that galaxy formation must be
inefficient in \lcdm\ (e.g., \citealt{fukugita2004}), meaning that only a
relatively small fraction of the baryonic allotment of a dark matter halo
($\sim 20\%$ for MW-mass haloes) will be converted into stars by $z=0$. Strong
feedback from galaxy formation can change the structure of dark matter haloes,
reducing their mass within a given radius compared to what would be obtained in a DMO
version (e.g., \citealt{Vogelsberger2014a, Schaller2015}). It is therefore of
great interest to study precisely how inferences about the MW's mass profile
change from using DMO simulations -- which, for given cosmological parameters,
are uniquely predicted -- to using cosmological hydrodynamic simulations.

\begin{figure}
\centering
\includegraphics[width=\linewidth]{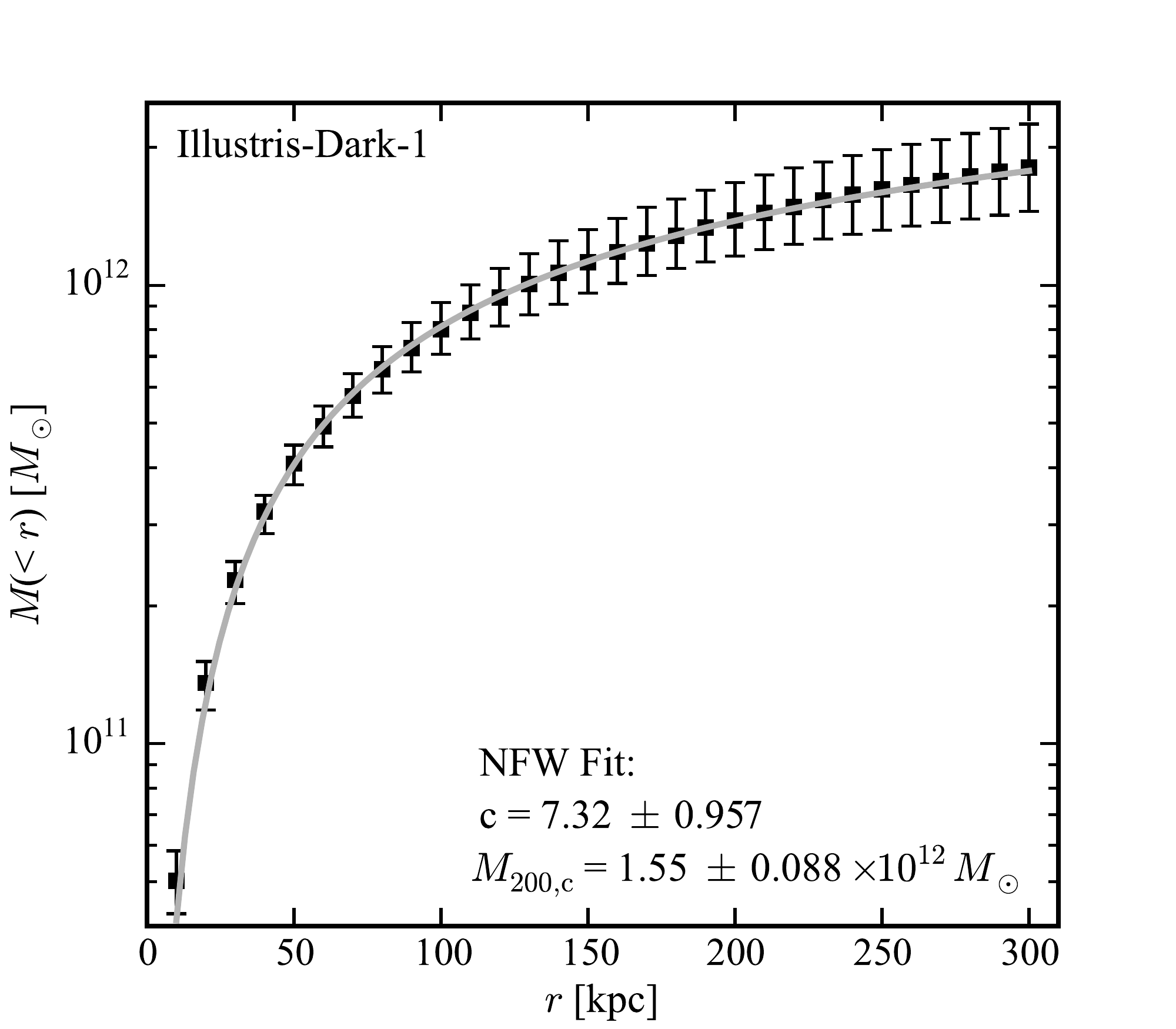}
\caption{The cumulative mass distribution from \illustris-Dark-1 (black points
  with error bars), along with the best-fitting NFW profile (grey line), derived
  assuming the D12 constraint on $\mfifty$. This figure can be directly
  compared to the left-hand panel of Fig.~\ref{mInR}, which shows the same
  quantities from the full hydrodynamic run. While both versions of Illustris
  are well-fitted by NFW profiles, the fit parameters differ substantially
  between the two: the DMO run is fitted by a higher-mass (37\% higher),
  lower-concentration (40\% lower) halo. If DMO runs are used for modelling the
  MW mass distribution based on $\mfifty$, or a similar constraint, this effect
  must be taken into account.}
\label{mInR-nfw}
\end{figure}

\begin{table}
\setlength{\extrarowheight}{5pt}
\caption{Median values, along with 68\% and 90\% confidence intervals, for a
  variety of mass measures explored in this paper (similar to
  Table~\ref{table:confidenceIntervals}); all values are in units of
  $10^{12}\,\msun$. In contrast to  
  Table~\ref{table:confidenceIntervals}, however, these results use the
  \illustris-Dark simulations.}
\resizebox{\linewidth}{!}{
\setlength{\extrarowheight}{5pt}
\begin{tabular}{ l| l l}
  \hline
  & D12 & G14\\
  \hline 
  \hline 
  & \multicolumn{2}{c}{Illustris-Dark-1}\\
  $M_{\rm 200,c}$ & $1.57^{+0.460\,(1.28)}_{-0.343\,(0.519)}$ & $0.836^{+0.296\,(0.605)}_{-0.220\,(0.336)}$ \\
  $M_{\rm vir}$ & $1.88^{+0.676\,(1.65)}_{-0.445\,(0.654)}$ & $0.993^{+0.382\,(0.805)}_{-0.275\,(0.417)}$ \\
  $M_{\rm 200,m}$ & $2.15^{+0.841\,(1.91)}_{-0.545\,(0.808)}$ & $1.12^{+0.466\,(0.997)}_{-0.318\,(0.482)}$ \\
  $M(< 100\, \kpc)$ & $0.803^{+0.116\,(0.198)}_{-0.095\,(0.161)}$  & $0.521^{+0.105\,(0.169)}_{-0.095\,(0.154)}$ \\
  $M(< 250\, \kpc)$ & $1.62^{+0.357\,(0.830)}_{-0.299\,(0.452)}$ & $0.971^{+0.283\,(0.548)}_{-0.231\,(0.355)}$\\
  \hline
  & \multicolumn{2}{c}{Illustris-Dark-2}\\
  $M_{\rm 200,c}$ & $1.59^{+0.515\,(1.48)}_{-0.344\,(0.531)}$ & $0.838^{+0.296\,(0.593)}_{-0.219\,(0.338)}$ \\
  $M_{\rm vir}$ & $1.91^{+0.712\,(1.86)}_{-0.452\,(0.681)}$ & $0.991^{+0.399\,(0.793)}_{-0.268\,(0.412)}$ \\
  $M_{\rm 200,m}$ & $2.18^{+0.920\,(2.19)}_{-0.565\,(0.835)}$ & $1.12^{+0.476\,(1.02)}_{-0.312\,(0.478)}$ \\
  $M(< 100\, \kpc)$ & $0.807^{+0.116\,(0.204)}_{-0.095\,(0.161)}$ & $0.523^{+0.103\,(0.170)}_{-0.094\,(0.153)}$ \\
  $M(< 250\, \kpc)$ & $1.63^{+0.396\,(0.968)}_{-0.290\,(0.462)}$ & $0.971^{+0.290\,(0.548)}_{-0.221\,(0.352)}$\\
  \hline
  & \multicolumn{2}{c}{Illustris-Dark-3}\\
  $M_{\rm 200,c}$ & $1.61^{+0.567\,(1.39)}_{-0.349\,(0.530)}$  & $0.859^{+0.302\,(0.665)}_{-0.231\,(0.348)}$\\ 
  $M_{\rm vir}$ & $1.96^{+0.760\,(1.98)}_{-0.488\,(0.708)}$ & $1.01^{+0.389\,(0.904)}_{-0.271\,(0.423)}$ \\ 
  $M_{\rm 200,m}$ & $2.23^{+1.00\,(2.31)}_{-0.601\,(0.868)}$ & $1.15^{+0.459\,(1.09)}_{-0.316\,(0.487)}$ \\ 
  $M(< 100\, \kpc)$ & $0.826^{+0.115\,(0.213)}_{-0.113\,(0.177)}$ & $0.526^{+0.107\,(0.178)}_{-0.090\,(0.152)}$ \\ 
  $M(< 250\, \kpc)$ & $1.65^{+0.417\,(0.901)}_{-0.311\,(0.481)}$ & $0.988^{+0.292\,(0.599)}_{-0.225\,(0.361)}$ \\ 
  \hline
  \hline
\end{tabular}
}

\label{table:confidenceIntervals_dmo}
\end{table}

The first test we perform to gauge the effects of including galaxy formation
physics on mass inferences is to rerun our analysis on the DMO versions of
\illustris.  Fig.~\ref{mInR-nfw} shows the results of applying the D12
constraint to \illustris-Dark-1. It can be directly compared to
Fig.~\ref{mInR}, in which the D12 constraint was applied to the hydrodynamic
version of \illustris-1. Relative to the full \illustris\ simulation, inferences
based on the DMO version result in a significantly higher estimate of
$M_{\rm 200,c}$ ($1.5\times 10^{12}$ versus $1.1\times 10^{12}\,\msun$) and a
significantly lower version of the NFW concentration ($c=7.4$ versus
12.3). Table~\ref{table:confidenceIntervals_dmo} provides an alternate version
of Table~\ref{table:confidenceIntervals} in which all constraints are obtained
using the DMO version of \illustris-1. In all cases, the net effect of using the
DMO run rather than the hydrodynamic version is to infer \textit{higher} values
for a given aperture mass.

We can use the \illustris\ suite to perform an additional test of the effects of
galaxy formation on the mass distribution within dark matter halos (and for
accompanying inferences on the mass distribution of the MW): since \illustris\
and \illustris-Dark share the same initial conditions, individual dark matter
halos can be matched between the two simulations (for details, see section~3.2
of \citealt{Vogelsberger2014b}). In this way, we can study the effects of galaxy
formation on a halo-by halo basis by identifying the DMO analogue of each halo in
the full \illustris\ run and comparing the resulting mass distributions.

\begin{figure*}
\centering
\includegraphics[width=0.49\linewidth]{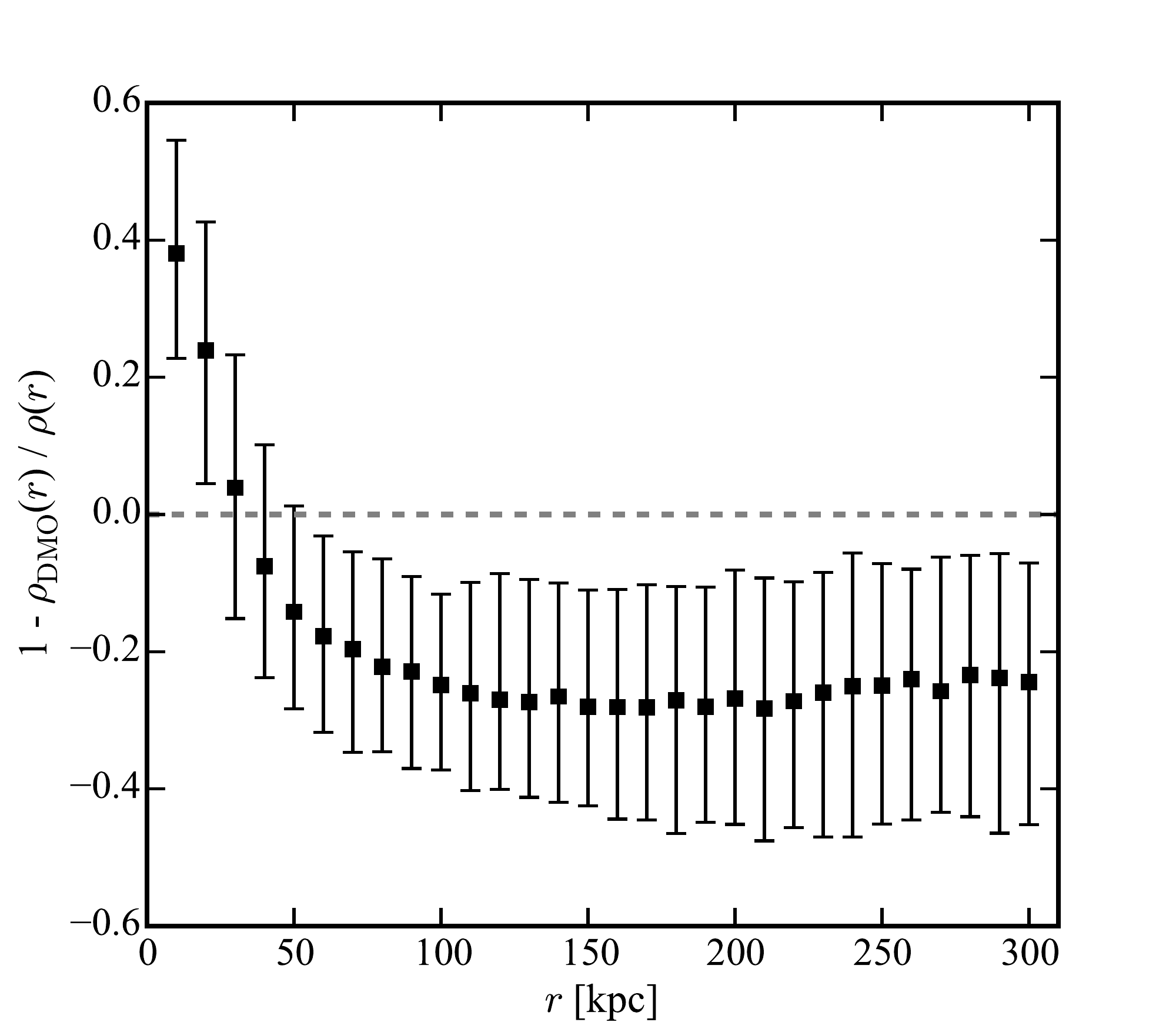}
\includegraphics[width=0.49\linewidth]{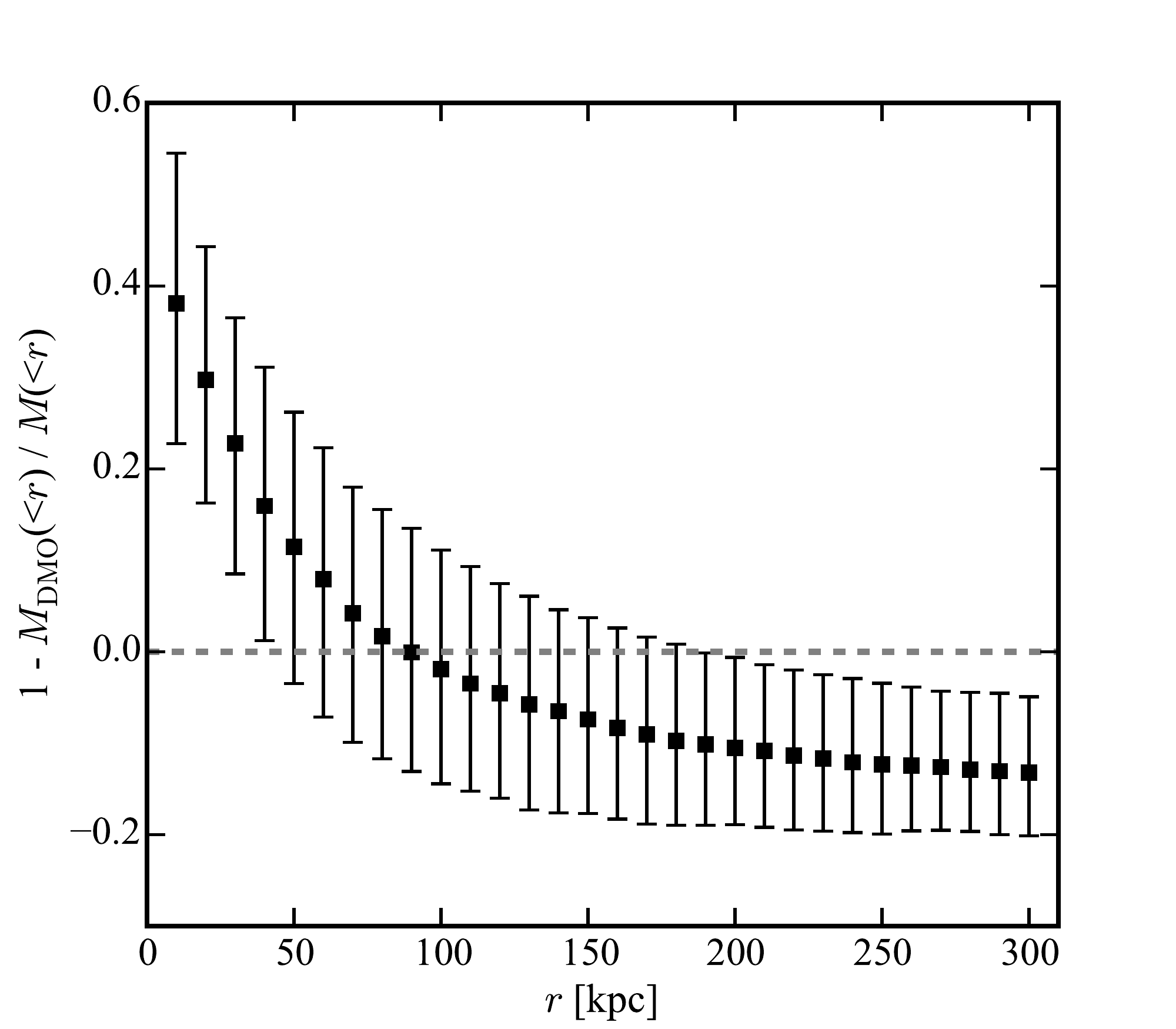}
\caption{Fractional differences in the density (left) and enclosed mass
  (right) profiles between \illustris-1 and \illustris-Dark-1, where haloes
  are individually matched across the two simulations (see the text for details). Data points represent the median differences between Illustris-1 and Illustris-Dark-1, while error bars show the central 68\% range of the data.  On small
  scales, the inclusion of baryonic physics results in more mass at a given radius owing to the
  formation of the central galaxy. On large scales, however, feedback causes an
  overall reduction in mass on a halo-by-halo basis for the full hydrodynamic
  simulation relative to the DMO run. The effect in the density profile is $\sim
  20\%$ at large radii, while the effect in the cumulative mass profile is $\sim
  10\%$ at large distances.}
\label{dark-v-hydro}
\end{figure*}

Fig.~\ref{dark-v-hydro} shows the results of this comparison, for which we use
haloes in \illustris-1 falling within the 68\% confidence interval of
$M_{\rm 200,c}$ computed using the D12 constraint (see
Table~\ref{table:confidenceIntervals}) -- assigning equal weight to all such
haloes -- and their counterparts in the DMO run. The left-hand panel shows how the
density profiles are affected at each radius. On small scales
($r \la 30\,\kpc$), the hydrodynamic run has higher densities on a halo-by-halo
basis. This is caused by the formation of the central galaxy, both through its
mass and through any adiabatic contraction. On larger scales
($r \ga 40 \,\kpc$), a given halo in the hydrodynamic run is less dense than its
equivalent in the DMO run by approximately 20\%. This reduction in density is
likely caused by outflows and the loss of gas mass (or the prevention of gas
accretion). The effect on the cumulative mass distribution is shown in the right-hand
panel of Fig.~\ref{dark-v-hydro}. On a halo-by-halo basis, the hydrodynamic
run results in larger masses out to $\approx 100$ kpc; on larger scales, the
masses in the DMO run are larger, with the difference reaching an asymptotic
value of $\approx 10\%$ at 250-300 kpc. As discussed in
Section~\ref{sec:discussion}, the details of the reduction in mass may depend on
the adopted models of galaxy formation modelling.

\subsection{The Stellar Mass of the Galaxy}

\begin{table}
\setlength{\tabcolsep}{15pt}
\setlength{\extrarowheight}{5pt}
\caption{Inferred values of $\mstar$, in units of $10^{10}\,\msun$, using the
  D12 (column 2) and G14 (column 3)
  constraints on $\mfifty$. The quoted errors are the 68\% and 90\% confidence
  intervals. The \illustris\ feedback prescriptions were calibrated for the
  highest-resolution simulation (\illustris-1), so perfect convergence in
  $\mstar$ across the three simulations is not expected.}  
\begin{tabular}{ l| l l}
	\hline
& D12 & G14\\
	\hline
  \rowcolor{LtGray}
	Illustris-1 & $5.04^{+1.47\,(2.72)}_{-1.32\,(2.03)}$ & $2.41^{+0.98 \,(1.74)}_{-0.72\, (1.12)}$ \\
	Illustris-2 & $4.09^{+1.19 \, (2.15)}_{-1.15\,(1.66)}$ & $1.83^{+0.76\,(1.36)}_{-0.59\, (0.90)}$ \\
	Illustris-3 & $2.57^{+0.79\,(1.52)}_{-0.71\,(1.07)}$ & $1.03^{+0.48\,(0.92)}_{-0.35\,(0.55)}$\\
	\hline
\end{tabular}
\label{table:stellarMass_mfifty}
\end{table}

We can also use the technique explored in the previous sections to compute the
galaxy stellar masses from \illustris\ that are consistent with the adopted mass
constraints at 50 kpc. Table~\ref{table:stellarMass_mfifty} gives the median
values as well as 68\% and 90\% confidence intervals based on the D12 and G14
constraints in each of the three \illustris\ resolution levels. Unlike
the total enclosed mass at large radii, which is well-converged across the three
different \illustris\ resolutions, the stellar masses in these haloes increase by
a factor of $\sim 2$ from \illustris-3 to \illustris-1. This difference is not
large enough to be reflected in stellar mass functions (which are reasonably
similar for the different resolution levels studied here; see, e.g.,
\citealt{Vogelsberger2013} and \citealt{Torrey2014}). It is larger than the
uncertainty on the measured $\mstar$ of the MW, however: most recent estimates
for the Galaxy fall in the range $\mstar = 5-6.5\times 10^{10}\,\msun$ (e.g.,
\citealt{McMillan2011, Bovy2013, Licquia2015}).\\[-0.3cm]

Differences in the simulated stellar masses at the factor of $\sim 2$ level are
unsurprising, as the galaxy formation models used in the \illustris\ suite were
calibrated at the resolution of \illustris-1; we would not expect the same
models to work identically at significantly lower resolution. Specifically, the
minimum resolution required for the feedback implementation in \illustris\ to
produce a realistic galaxy population is not achieved in \illustris-3 \citep{Vogelsberger2013}. We
therefore consider the results from \illustris-1 to be the most reasonable
comparison to make with observations.

We adopt the measurement of \citet[hereafter, LN15]{Licquia2015}, in
  which the authors used results derived in \cite{Bovy2013} to obtain
$\mstar=6.08 \pm 1.14\times 10^{10}\,\msun$, as a representative value of the
stellar mass of the MW and use it as a reference point in what follows.
Comparing this number to the results for \illustris-1 in
Table~\ref{table:stellarMass_mfifty}, we see that D12 agrees well with the
observed value, while G14 is substantially lower. This is not surprising, given
the results of Table~\ref{table:confidenceIntervals}. The very low value of
$M_{\rm 200,c}$ obtained based on G14 is much lower than the typical value found
for haloes with the stellar mass of the MW via either abundance matching
\citep{guo2010, behroozi2013c, moster2013}, galaxy-galaxy lensing
\citep{Mandelbaum2016}, or satellite kinematics (e.g., \citealt{watkins2010,
  Boylan-Kolchin2013}). Even accounting for possible differences in halo masses
of red and blue galaxies at fixed stellar mass \citep{Mandelbaum2016}, the MW
would be a strong outlier if its mass is as low as the median value indicated by
our analysis using the G14 constraint on $\mfifty$.

As noted in Section~\ref{sec:methods}, our methodology for constraining the mass
profile of the MW is quite general. While we have focused on constraining the
total mass at large radii based on measurements of the total mass within 50 kpc,
we can instead use other quantities -- for example, $\mstar$ -- for our
inference. Following the same procedure outlined in Section~\ref{sec:methods},
we estimate $\mfifty$, $\mhundred$, and the three spherical overdensity masses
used above based on LN15's determination of $\mstar$; the results are presented
in the second column of Table~\ref{table:mstar}.  The results are very similar
to those obtained using the D12 determination of $\mfifty$, with LN15-based
estimates being 5-7\% higher (the results are approximately a factor of 1.6--2
larger than G14-based estimates).

\begin{figure}
\centering
\includegraphics[width=1.05\linewidth]{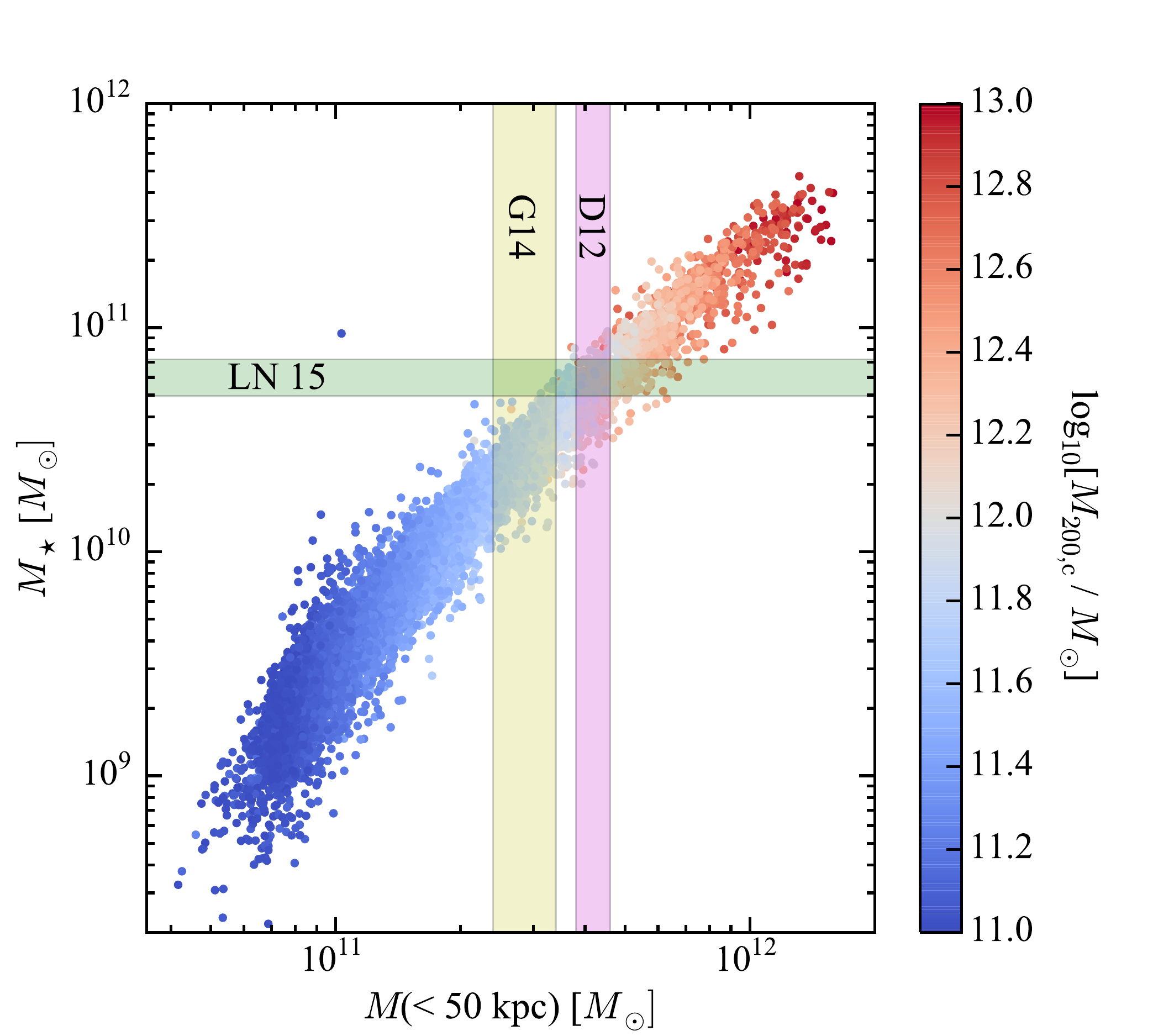}
\caption{The correlation between $\mstar$ and $\mfifty$ for all haloes in the
  Illustris-1 sample; the haloes are coloured by the value of $M_{\rm 200,c}$.
  Vertical shaded bands show G14 (yellow) and D12 (magenta) determinations of
  $\mfifty$ while the horizontal band shows the LN15
  determination of $\mstar$ for the MW.  Very few haloes agree with both
  the G14 measurement of the total mass at 50 kpc and the MW's stellar
  mass; many more of the simulated galaxies match the D12 value for $\mfifty$
  and the LN14 $\mstar$ value simultaneously.}  
\label{mstar-m50}
\end{figure}

Since $\mstar$ and $\mfifty$ can be considered independent variables, we can
also study the joint probability of obtaining various mass measures conditioned
on $\mstar$ and $\mfifty$. These joint constraints, using D12's estimate of
$\mfifty$, are given in the third column of Table~\ref{table:mstar}. The joint
constraints are similar to both the estimates using $\mstar$ alone and the
estimate using $\mfifty$ (from D12) alone, which is a result of the good
agreement of each of these estimates individually. Had we used the G14 value of
$\mfifty$, the constraints would have shifted substantially. This is highlighted
in Fig.~\ref{mstar-m50}, which shows the \illustris-1 data in $\mstar-\mfifty$
space; each halo assigned a colour according to its value of $M_{\rm 200,c}$. The
intersection of the D12 and LN15 constraints falls along the main locus of the
points while the G14 constraint intersects the LN 15 constraint in a part of
parameter space with very few haloes. The D12 and LN15 measurements are therefore
in good agreement based on the \illustris\ haloes, while the G14 and LN15
measurements are not.

\begin{table}
\caption{The 68\% and 90\% confidence intervals of various mass measures of the
  MW (all in units of $10^{12}\,\msun$) inferred using the LN15
  measurement of the MW's $\mstar$ alone (column 2) and jointly with the
  D12 constraint on $\mfifty$ (column 3).}
\resizebox{\linewidth}{!}{
\setlength{\extrarowheight}{5pt}
\begin{tabular}{l l l}
  \hline
  & $P(M | \mstar)$ & $P(M|\mstar, M_{\rm D12})$\\
  \hline
  $M_{\rm 200,c}$ & $1.19^{+0.62\,(1.44)}_{-0.35\,(0.52)}$ &
                                                             $1.13^{+0.36\,(0.80)}_{-0.22\,(0.36)}$\\ 
  $M_{\rm vir}$ & $1.40^{+0.81\,(1.86)}_{-0.43\,(0.64)}$ & $1.33^{+0.49
                                                           \, (1.20)}_{-0.29\,(0.46)}$\\
  $M_{\rm 200,m}$ & $1.60^{+1.00\,(2.15)}_{-0.53\,(0.78)}$ &
                                                                 $1.50^{+0.65\,(1.66)}_{-0.35\,(0.54)}$ \\ 
  $M(<100\, \kpc)$ & $0.725^{+0.179\,(0.358)}_{-0.145\,(0.231)}$ &
                                                                        $0.707^{+0.081\,(0.155)}_{-0.087\,(0.144)}$ \\ 
  $M(<250\, \kpc)$ & $ 1.29^{+0.51\,(1.08)}_{-0.34\,(0.51)}$ &
                                                                   $1.23^{+0.34\,(0.67)}_{-0.21\,(0.36)}$\\
\hline
\end{tabular}
}
\label{table:mstar}
\end{table}

\section{Discussion and future prospects}
\label{sec:discussion}
\begin{figure}
\centering
\includegraphics[width=\linewidth]{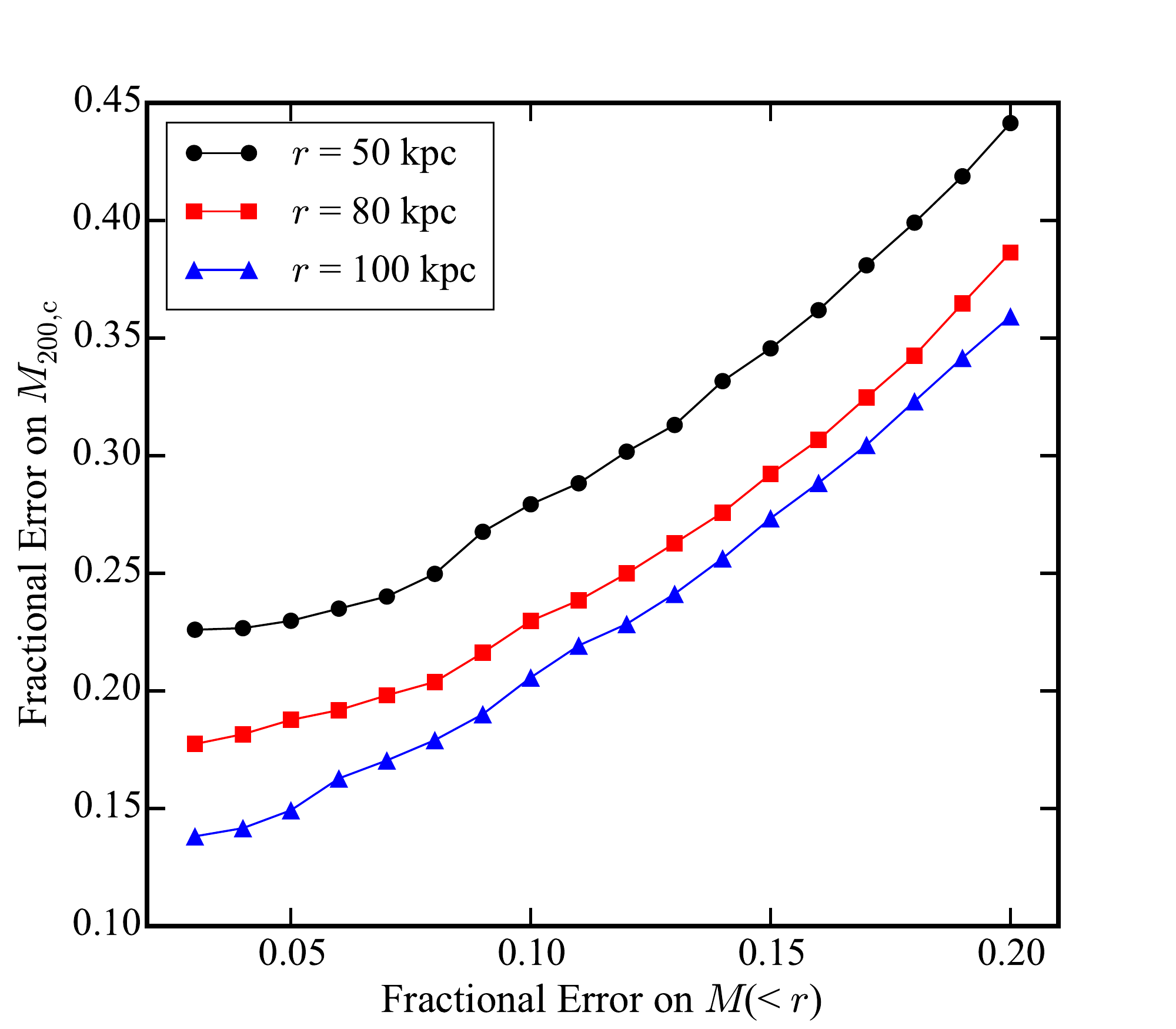}
\caption{The precision attained in measurements of $M_{\rm 200,c}$ as a function
  of the precision in the input constraint. We consider input constraints of the total mass within 50, 80, and 100 kpc (black circles, red squares, and
  blue triangles, respectively) to show how more precise determinations of
  masses within larger radii can affect the inferred value of $M_{\rm 200,c}$. The
  figure shows the trade-off between precision and distance: for example, an
  error of 15\% in $\mhundred$ results in the same precision in the estimate of
  $M_{\rm 200,c}$ as an error of 9\% in $\mfifty$. }
\label{fig:errorsVsRad}
\end{figure}
As larger samples of halo stars at greater distances become available, it may
become possible to constrain the mass of the MW enclosed within 80 or even 100 kpc (see,
e.g., \citealt{gnedin2010, Cohen2015} for initial work in this direction). Such
measurements would have the benefit of providing stronger constraints on the
virial mass of the MW. Fig.~\ref{fig:errorsVsRad} shows the fractional
uncertainty in $M_{\rm 200,c}$ as a function of the error in the mass contained
within 50 (black circles), 80 (red squares), and 100 kpc (blue triangles). At a
fixed uncertainty in $M(<r)$, the implied uncertainty in $M_{\rm 200,c}$ does
indeed become smaller as one moves to greater Galactocentric distance.

The figure quantifies how improving uncertainties at a given distance will be
reflected in uncertainties on $M_{\rm 200,c}$: for example, reducing the error
on $\mfifty$ from 10 to 5\% would reduce the error on $M_{\rm 200,c}$ from 28 to
23\%. On the other hand, an measurement of the mass within 80 kpc that is
accurate to 10\% results in an error of 22\% in $M_{\rm 200,c}$, while the same
accuracy on a measurement of the mass within 100 kpc of the Galaxy would yield
errors of 19\% in $M_{\rm 200,c}$. The figure also shows the fundamental
limitations in extrapolating to $M_{\rm 200,c}$ based on measured aperture
masses within smaller radii. Some level of irreducible uncertainty is unavoidable in standard cosmological models, as extrapolation from mass at a given radius to the virial radius depends on the halo concentration (e.g. \citealt{Navarro1997, Bullock2001}). For example, consider the recent study of
\citet{Williams2015}, who found that
$\mfifty=4.48^{+0.15}_{-0.14}\times 10^{11}\,\msun$, or an error of
approximately 3\% on $\mfifty$. Using this constraint, we obtain
$M_{\rm 200,c}=1.25^{+0.35}_{-0.18}\times10^{12}\,\msun$; the uncertainty on the
derived value of $M_{\rm 200,c}$ remains large in spite of the high precision of
the input measurement. Fig.~\ref{fig:errorsVsRad} makes it clear that
measurements of the mass within 50 (80, 100) kpc will result in an uncertainty
on $M_{\rm 200,c}$ of no better than 23\% (17\%, 14\%).

A central assumption of the techniques we employ here is that \illustris-1
provides a faithful representation of galaxies and the effects of galaxy
formation on dark matter halo structure. Since cosmological hydrodynamic
simulations are still at the point of relying on subgrid models of physics, and
will be for the foreseeable future, a logical extension of our work would be to
investigate predictions in future generations of simulations to test the
robustness of our results. It would also be interesting to compare the results
we have obtained with \illustris\ to the Eagle simulations, as the galaxy
formation modelling employed there is somewhat different. Given the differences
seen in the ratio of masses in hydrodynamic to DMO simulations in \illustris\
versus Eagle (compare fig.~7 of \citealt{Vogelsberger2014b} and 
fig.~1 of \citealt{Schaller2015}), such a comparison would be timely.

One effect that appears to be particularly important for setting the amount of
mass reduction for a given halo in the hydrodynamic run relative to its
counterpart in the DMO version is the underlying model of AGN
feedback. \citet{Vogelsberger2014a} adopted an AGN model that drives
\textit{very} strong outflows, perhaps unrealistically so
\citep{Genel2014}. Forthcoming updates to the \illustris\ suite will use
modified versions of AGN feedback that are less powerful and may result in
different modifications of the large-scale halo properties of galaxies, which
may in turn affect how $\mfifty$ maps on to $M_{\rm 200,c}$.

To explore the potential impact of this effect on our results, we use the
current generation of \illustris\ and compare the effects of BH mass for
galaxies of a fixed halo mass (we use the haloes that are closest to the median
value of $M_{\rm 200,c}$ found in \illustris-Dark-1 using the D12
constraint). We rank this sample according to black hole mass and then compute
the difference in mass in the hydrodynamic simulation relative to the DMO
run. There is indeed a difference: the galaxies with the highest-mass black
holes show a 20\% reduction in their overall mass, on average, while the galaxies
with the lowest mass black holes see a 10\% reduction in mass compared to their DMO counterparts. The total halo mass therefore appears to depend somewhat on the choice of black hole feedback model, although this does not appear to be a large source of uncertainty in
our predictions. Future generations of \illustris-like simulations with modified
black hole feedback models will allow us to directly test the effects on
inferences regarding the MW mass. 

It is not entirely obvious how the effects of vigorous feedback propagate
through our analysis, as this will depend on the change in enclosed mass within 50 kpc
relative to the change in enclosed mass within larger radii. However, given that the black
hole feedback in the current version of \illustris\ may be too effective and
that the larger-mass black holes correlate with larger reductions in halo mass as compared to lower-mass black holes, it is likely that any modified prescriptions will result in
slightly higher inferences on the total halo mass compared to our current
results, should there be a difference.

Future work would also benefit significantly from cosmological simulations with
larger volumes and higher mass resolution. Importance sampling
relies on having a well-sampled parameter space, which can be an issue if not
many haloes match the desired constraint(s) (see \citealt{Busha2011} and
\citealt{Gonzalez2014} for more details). Our current analysis has many haloes
contributing significant weights: 870 and 2196 haloes contribute weights that are
at least 10\% of the maximum possible weight
($W_{\rm max}=1/\sqrt{2\rm{\pi}\sigma^2}$ from equation.~\ref{eq:singleprob}) for the D12
and G14 constraints, respectively. However, if we wish to add additional
restrictions -- based on morphology, disc size, star formation history, or
specific star formation rate, for example -- the sample would likely become
significantly smaller, which would be the limiting factor in the conclusions we
could draw. With larger sample sizes, such concerns would be eliminated. From
Fig.~\ref{mstar-m50}, joint constraints on $\mfifty$ and $\mstar$ are unlikely
to be strongly affected by sample size unless a much larger volume produced many
haloes with much larger stellar masses at fixed halo mass [in which case, the G14
measurement of $\mfifty$ would be more consistent with the simulation results
than it is at present].

\section{Conclusions}
\label{sec:conclusions}

In this paper, we have explored how the \illustris\ suite can be used to inform
our understanding of the mass distribution around the MW. Our main
conclusions are as follows.
\begin{itemize}
\item The mass profiles of haloes consistent with a given constraint on $\mfifty$
  differ substantially between DMO and hydrodynamic versions of \illustris.\ Using DMO
  simulations to extrapolate from 50 kpc to larger radii results in an
  overestimate of the halo mass and an underestimate of the halo concentration.
\item The effects of baryonic physics on the mass distribution of MW-like
  systems in \illustris\ are substantial: by matching haloes between the DMO and
  hydrodynamic simulations, we find that the latter have more mass on small
  scales and less mass on large scales. The asymptotic difference in the total
  mass density at large radii is approximately 20\%.
\item Since different feedback models result in very different effects on the
  mass distribution of dark matter even at large distances from halo centres
  (e.g., fig.~7 of \citealt{Vogelsberger2014b} compared to fig.~1 of
  \citealt{Schaller2015}), it is imperative to test how inferences on the
  mass of the MW depend on galaxy formation modelling.
\item The mass distribution in the inner $\sim 20$ kpc is not converged in the
  \illustris\ suite [see \citealt{Schaller2016} for similar results in the Eagle
  simulations]; this is a much larger distance than the formal convergence
  radius for the dark matter simulations. Results regarding the density
  distribution for $r \la 20$ kpc must therefore be interpreted with caution,
  and our best-fitting NFW profiles for the hydrodynamic simulations, which were
  obtained over the radial range of 40-300 kpc, should not
  be extrapolated to smaller radii.
\item The relationship between $\mfifty$ and $M_{\rm 200,c}$ in \illustris-1 is
  well-described by a log-quadratic relationship
  (equation.~\ref{eq:mfifty_fit}). This relationship enables the translation of any
  existing or future constraint on $\mfifty$ into a measurement
  $M_{\rm 200, c}$.
\item The constraints on $\mfifty$ derived by D12
  ($4.2 \pm 0.4\times 10^{11}\,\msun$) and G14
  ($2.9 \pm 0.4 \times 10^{11}\,\msun$) predict very different values for the
  virial mass of the Galaxy's halo when using \illustris: for D12, we find
  $M_{\rm 200,c}=1.12^{+0.37}_{-0.24}\times 10^{12}\,\msun$ (68\% confidence),
  while for G14, we find
  $M_{\rm 200,c}=0.612^{+0.196}_{-0.148}\times10^{12}\,\msun$ (68\%
  confidence). The values for $\mvir$ and $M_{\rm 200,m}$ are 17\% and 32\%
  larger, respectively.
\item \illustris\ haloes that have galaxies with stellar masses consistent with
  measurements of the MW's $\mstar$ have significantly more mass within
  $50\,\kpc$ than the result of G14; the measurements of D12 and
  \citet{Williams2015} are in much better agreement with \illustris\ haloes that
  match the observed value of $\mstar$. In particular, almost no haloes in
  \illustris\ jointly satisfy the G14 constraint and the LN15 measurement of
  $\mstar$ for the MW.
\item From our analysis of the \illustris\ simulation, even an infinitely precise measurement of $\mfifty$ would result in an
  uncertainty of $>$20\% in $M_{\rm 200,c}$. The same uncertainty can be achieved for 10\% errors on $M(<80\,\kpc)$ or 12\% errors on $\mhundred$. A
  measurement of $\mhundred$ that is accurate to 5\% will translate into 15\%
  uncertainties on $M_{\rm 200,c}$.
\end{itemize}

As ever larger and ever more realistic hydrodynamic simulations become
available, so too will better statistical constraints on the mass profile of our
Galaxy.

\vspace{0.2cm}

\section*{Acknowledgements} 
We thank Joss Bland-Hawthorn, Nitya Kallivayalil, Julio Navarro, and Annalisa
Pillepich for helpful conversations. The analysis of the \illustris\ data sets
for this paper was done using the Odyssey cluster, which is supported by the FAS
Division of Science, Research Computing Group at Harvard University. MB-K
acknowledges support provided by NASA through a \textit{Hubble Space Telescope} theory
grant (programme AR-12836) from the Space Telescope Science Institute (STScI),
which is operated by the Association of Universities for Research in Astronomy
(AURA), Inc., under NASA contract NAS5-26555. PT acknowledges support from NASA
ATP Grant NNX14AH35G. LH acknowledges support from NASA grant NNX12AC67G and NSF
grant AST-1312095.

\bibliography{ms_clean}

\label{lastpage}
\end{document}